\documentclass[aps,showpacs,11pt]{revtex4}
\usepackage{dcolumn}
\usepackage{graphicx}
\usepackage{amsmath}
\usepackage{amsfonts}
\usepackage{amssymb}
\usepackage{psfrag}
\usepackage{wrapfig}
\usepackage{subfigure}
\usepackage{makeidx}
\usepackage{bm}
\usepackage{epsf}
\usepackage{hyperref}
\usepackage{color}
\usepackage{float}
\newcommand{\no}{\nonumber}
\def\/{\over}

\begin{document}

\title{Stability of Reissner-Nordstr\"{o}m black hole in de Sitter background under charged scalar perturbation}

\author{Zhiying Zhu$^{1,2}$, Shao-Jun Zhang$^{1}$, C. E. Pellicer$^{3}$, Bin Wang$^{1}$, Elcio Abdalla$^{4}$}
\affiliation{ $^1$ Department of Physics and Astronomy,
Shanghai Jiao Tong University, Shanghai 200240, China\\
 $^2$ Department of Physics and Electronic Science, Changsha
University of Science and Technology, Changsha, Hunan 410076, China\\
$^3$ Escola de Ci\^{e}ncias e Tecnologia, Universidade Federal do Rio Grande do Norte, Caixa Postal 1524, 59072-970, Natal, Rio Grande do Norte, Brazil\\
$^4$ Instituto de F\'{i}sica, Universidade de S\~{a}o Paulo, CEP 05315-970, S\~{a}o Paulo, Brazil}

\begin{abstract}

We find a new instability in the four-dimensional
Reissner-Nordstr\"{o}m-de Sitter black holes  against
charged scalar perturbations with vanishing
angular momentum, $l=0$. We show that such an
instability is caused by superradiance. The
instability does not occur for a larger angular
index, as explicitly proved for $l=1$. Our
results are obtained from a numerical
investigation of the time domain-profiles of the
perturbations.

\end{abstract}

\pacs{04.70.Bw, 42.50.Nn, 04.30.Nk}


\maketitle

\section{introduction}
Perturbations around black holes have been intriguing objects of discussions for decades.  One of the main reasons is their
astrophysical interest.  Real black holes always have interactions with the respective astrophysical environment through
absorption or evaporation processes.  Starting from  analyses of the perturbations around black holes, we can inquire about
their stability.  If  black holes are unstable under small perturbations, they will inevitably be destabilized and disappear or transform
dynamically into another object, thus they simply cannot exist in nature as it is originally defined. For the (3+1)-dimensional
asymptotically flat black holes,  such as Schwarzschild, Reissner-Nordstr\"{o}m (RN)  and Kerr black holes, the corresponding
stability has been checked thoroughly against different kinds of perturbations including neutral scalar, electromagnetic and
gravitational perturbations and they were found stable. Analyses of these perturbations and proofs of stability of
(3+1)-dimensional Schwarzschild-de Sitter (dS), RN-dS, and Kerr-dS black holes have been reported.  Recently, motivated by the
discovery of the correspondence between physics in the anti-de Sitter (AdS) spacetime and conformal field theory (CFT) on its
boundary (AdS/CFT), the perturbations around four-dimensional AdS black holes  have been examined and (3+1)-dimensional AdS black
holes were found stable under neutral scalar, electromagnetic and gravitational perturbations.  It was concluded that all of
the considered four-dimensional black holes tested for stability are stable, except the string theory generalization of Kerr-Newman black holes
whose stabilities have not been tested  due to the difficulty in decoupling the angular
variables in their perturbation equations. For a review on this topic, see \cite{Konoplya:2011, Wang:2005} and references therein.

The physics in higher dimensions is much richer. In contrast to the four-dimensional results, various instabilities have
been found.  In a wide class of $D \geq 4$ configurations, such as black strings and black branes, the Gregory-Laflamme
instability against linear perturbations was disclosed \cite{Konoplya:2011:295,Konoplya:2011:296}.  For high-dimensional black
holes in the Einstein-Gauss-Bonnet theory, it was found that instabilities occur at higher multipoles $l$, while the first few
lowest multipoles are stable \cite{Konoplya:2011:140,Konoplya:2011:299}.  Recently, in the high-dimensional RN-dS black hole
background, numerical investigations have uncovered the surprising result that the RN black holes in dS backgrounds are
unstable for spacetime dimensions larger than six \cite{Konoplya:2009, Cardoso:2010}.

The dynamical perturbations of a black hole background can usually be described by a single wavelike equation, where a
growing mode of the perturbation indicates the instability of the black hole. Besides such a procedure, there also exists a
superradiant instability in the black hole spacetime.  Considering the classical scattering problem for a perturbation
field in a black hole background, we can have superradiance, a phenomenon where the reflected wave has larger amplitude than
the incident wave.  If the effective potential contains an extra local minimum out of the black hole in addition to
the local maximum,  the superradiance will get successively amplified in the valley of the local potential near the black hole,
which will result in the superradiant instability destabilizing the black hole.  Reviews of the superradiance stability can be
found in Refs.~ \cite{Konoplya:2011, Cardoso:2013:arXiv:1307.0038}.   Most investigations on the superradiant instability were
concentrated on the rotating black holes.  Recently, the discussions of the superradiant stability were extended to the
charged black holes with charged scalar field perturbations \cite{Hod:2012,Hod:2013}. It was found that the superradiant
instability can happen more efficiently when one considers charged black holes and charged scalar fields
\cite{Cardoso:2013:50,Hod:2013-2,Zhang:2013}. It would be of great interest to extend the study of the superradiant stability to
black holes in the real Universe, namely the black hole in the dS backgrounds.

We will study small charged scalar field perturbations in the vicinity of a (3+1)-dimensional RN black hole in a de Sitter
background by doing numerical calculations.  The stability of such black holes under charged scalar perturbations has never
been examined previously. In the AdS spacetime, it was observed that the (3+1)-dimensional RN-AdS black hole can become
unstable due to the condensation of the charged scalar hair onto the black hole and finally the AdS black hole can be destabilized
under a small charged scalar field perturbation \cite{arXiv:1002.2679, Abdalla:2010,1111.6729}.  Does this observed instability
occur only for the AdS black holes because of their special spacetime properties?  Will the dynamical stability also happen
in other four-dimensional black hole backgrounds?  It is of great interest to test whether the (3+1)-dimensional RN-dS black
hole can remain stable as  other four-dimensional black holes  under such charged scalar perturbing fields. This is the first
motivation of the present paper. Besides, we would like to explore whether the stability against linearized dynamical
perturbations relates to the superradiant stability due to the superradiant amplification of charged scalar waves.

The organization of the paper is as follows.  In Sec. II we review the (3+1)-dimensional RN-dS black hole background and
give the radial wave equation for a charged scalar field perturbation.
In Sec. III we  use the finite difference method to study the
time-domain perturbation evolution of the charged scalar field and test the stability against such perturbations.
In Sec. IV, we  derive the superradiant condition for the four-dimensional RN-dS black hole under a charged scalar perturbation, and examine
the relation between the dynamic stability and the superradiant stability. Finally, we  provide conclusions in Sec. IV.
We  use natural units with $G=\hbar=c=1$.

\section{METRIC, perturbation FIELDS AND EFFECTIVE
POTENTIALS}

We consider charged massive scalar
field perturbation in the RN-dS black hole
background with the metric
\begin{eqnarray}
ds^2=-f(r)dt^2+{1\/f(r)}dr^2+r^2(d\theta^2+\sin^2\theta d\phi^2)\;,
\end{eqnarray}
where
\begin{eqnarray}
f(r)=1-{2M\/r}+{Q^2\/r^2}-{\Lambda r^2\/3}\;.
\end{eqnarray}
The integration constants $M$ and $Q$ are the
black hole mass and electric charge,
respectively. $\Lambda$ is the positive
cosmological constant.

The spacetime causal structure depends strongly
on the zeros of $f(r)$. Depending on the
parameters $M, Q,$ and $\Lambda$, the function
$f(r)$ can have one to three or even no real
positive zeros (for a negative cosmological constant). For the RN-dS cases we are
interested in, $f(r)$ has three real, positive
roots $(r_c, r_+, r_-.)$, and a real and negative
root $r_0 = -(r_- + r_+ + r_c)$. The horizons
$r_-, r_+, r_c$ are denoted as black hole Cauchy, event, and
cosmological horizons respectively, which satisfy $r_- < r_+ <
r_c$.

The metric function $f(r)$ can be expressed as,
\begin{eqnarray}
f(r)={\Lambda\/3r^2}(r-r_+)(r_c-r)(r-r_-)(r-r_o)\;.
\end{eqnarray}
Introducing the surface gravity
$\kappa_i={1\/2}|df/dr|_{r=r_i}$ associated with
the horizon $r=r_i$, we can write the inverse of
the metric function
\begin{eqnarray}
{1\/f}=-{1\/2\kappa_-(r-r_-)}+{1\/2\kappa_+(r-r_+)}+{1\/2\kappa_c(r_c-r)}+{1\/2\kappa_o(r-r_o)}\;.
\end{eqnarray}
Thus we can obtain the analytic form of the
tortoise coordinate by calculating $r_*=\int
f^{-1}(r)dr$,
\begin{eqnarray}
r_*=-{1\/2\kappa_-}\ln\bigg({r\/r_-}-1\bigg)+{1\/2\kappa_+}\ln\bigg({r\/r_+}-1\bigg)-{1\/2\kappa_c}
\ln\bigg(1-{r\/r_c}\bigg)+{1\/2\kappa_o}\ln\bigg(1-{r\/r_o}\bigg)\;.
\end{eqnarray}
It is easy to see that $r_*\rightarrow -\infty$
as $r\rightarrow r_+$ and $r_*\rightarrow \infty$
as $r\rightarrow r_c$.

Consider now a charged scalar perturbation field
$\psi$ obeying the Klein-Gordon equation
\begin{eqnarray}
[(\nabla^\nu-iqA^\nu)(\nabla_\nu-iqA_\nu)-\mu^2]\psi=0\;,
\end{eqnarray}
where $q$ and $\mu$ are, respectively, the charge and mass of
the perturbing field and $A_\mu=-\delta^0_\mu Q/r$ denotes the electromagnetic potential of the black hole. By adopting the
usual separation of variables in terms of a
radial field and a spherical harmonic
$\psi_{lm}(t,r,\theta,\phi)=\Sigma_{lm}{1\/r}\Psi_{lm}(t,r)S_{lm}(\theta)e^{im\phi}$, the
Schr\"odinger-type equations in the tortoise
coordinate for each value of $l$ read
\begin{eqnarray}
-{\partial^2\Psi\/\partial t^2}+{\partial^2\Psi\/\partial r_*^2}+2iq\Phi{\partial\Psi\/\partial t}-V(r)\Psi=0\;,\label{radial eq}
\end{eqnarray}
where  $\Phi=-Q/r$ and the effective potential $V$ is given by
\begin{eqnarray}
V(r)=-q^2\Phi^2(r)+f(r)\bigg({l(l+1)\/{r^2}}+\mu^2+{f'(r)\/r}\bigg)\;.\label{V}
\end{eqnarray}

For a neutral massless scalar field perturbation,
the effective potentials were described
in Refs.~\cite{Brady:1997, Brady:1999, Molina:2004},
which vanish at the black hole and cosmological
event horizons. However, for the charged scalar
field perturbation, it is easy to see that at both
the black hole and cosmological event
horizons, the effective potentials are negative.
It is believed that the effective potential $V$
describes the scattering of $\psi$ by the
background curvature \cite{Ching:1995PRL}.
Usually, if the effective potential is negative
in some region, growing perturbation, can appear
in the spectrum, indicating an instability of the
system under such perturbations. However, in
Ref.~\cite{Bronnikov:2012}  it was argued that this is
not always true. They found that some potentials
with a negative gap still do not imply
instability. The criterion to determine whether a
system is stable or not against linear
perturbation is whether the time-domain profile
for the evolution of the perturbation is decaying
or not, or, in more general terms, the potential
has to permit the existence of bound states
\cite{bachelotbachelot}. Thus, in order to study
the stability of the RN-dS black hole against charged
scalar perturbation, we have to examine the
evolution of the perturbation. In calculating the
wave equation to obtain the evolution of the
perturbation, usually the quasinormal boundary
conditions should be employed by defining the
solution with the purely ingoing wave at the
black hole event horizon and outgoing wave at the
cosmological horizon.

\section{stability analysis}
We do not have analytic solutions to the
time-dependent wave equation with the effective
potentials considered here. We thus
discretize the wave equation (\ref{radial eq}).
Because of the appearance of the term
$2iq\Phi{\partial\Psi\/\partial t}$, the
procedure used in Ref.~\cite{Wang:2004} is not
convenient here.  One simple efficient
discretization, used for example in
Ref.~\cite{Abdalla:2010}, is to define
$\Psi(r_*,t)=\Psi(j\Delta r_*, i\Delta
t)=\Psi_{j,i}$, $V(r(r_*))=V(j\Delta r_*)=V_j$
and $\Phi(r(r_*))=\Phi (j\Delta r_*)=\Phi_j$, and to
write (\ref{radial eq}) as
\begin{eqnarray}
&&-{(\Psi_{j,i+1}-2\Psi_{j,i}+\Psi_{j,i-1})\/\Delta
t^2}+2iq\Phi_j
{(\Psi_{j,i+1}-\Psi_{j,i-1})\/2\Delta t} \no \\
&&\qquad \qquad \qquad +{(\Psi_{j+1,i}-2\Psi_{j,i}+\Psi_{j-1,i})\/\Delta
r_*^2}-V_j\Psi_{j,i}+O(\Delta t^2)+O(\Delta
r_*^2)=0\quad .
\end{eqnarray}
With the initial Gaussian distribution
$\Psi(r_*,t=0)=\exp[-{(r_*-a)^2\/2b^2}]$
and $\Psi(r_*,t<0)=0$,
we can derive the evolution of $\Psi$ by
\begin{eqnarray}
\Psi_{j,i+1}=-{{(1+iq\Phi_j\Delta t)\Psi_{j,i-1}}\/{1-iq\Phi_j\Delta t}}+{{\Delta t^2}\/{\Delta r_*^2}}
{{\Psi_{j+1,i}+\Psi_{j-1,i}}\/{1-iq\Phi_j\Delta t}}+\bigg(2-2{{\Delta t^2}\/{\Delta r_*^2}}-\Delta t^2V_j\bigg)
{\Psi_{j,i}\/{1-iq\Phi_j\Delta t}}\;.
\end{eqnarray}
In the following, we choose the parameters $a=10$
and $b=3$ in the Gaussian profile. Since the von
Neumann stability conditions usually require that
${{\Delta t}\/{\Delta r_*}}<1$, we use ${{\Delta
t}\/{\Delta r_*}}=0.5$ here. We will discuss the
behavior of the field $\varphi=\Psi/r$ in the
following.

It is necessary to point out that the time-domain
profiles of perturbations include  contributions
from all modes. It was argued that this method is
based on the scattering of the Gaussian wave on
the potential barrier and is independent of the
boundary conditions at the black hole event and
cosmological horizons \cite{Konoplya:2013}.
Therefore it includes all possible instabilities
due to different boundary conditions.

In order to extract dominant frequency from the
time-domain profile of the perturbation, we will
use the Prony method. We can fit the  profile
data by superposition of damping exponents
\cite{Berti:2007}
\begin{eqnarray}
\Psi(r,t)\simeq\sum_{i=1}^pC_ie^{-i\omega_i(t-t_0)}\;.\label{time-domain}
\end{eqnarray}
We consider a late time period, which starts at $t_0$ and ends at $t=N\Delta t+t_0$, where $N$ is an integer and $N\geq2p-1$.
Then Eq.~(\ref{time-domain}) is valid for each value from the profile data:
\begin{eqnarray}
x_n\equiv\Psi(r,n\Delta t+t_0)=\sum_{j=1}^pC_je^{-i\omega_jn\Delta t}=\sum_{j=1}^pC_jz_j^n\;.
\end{eqnarray}
The Prony method allows us to find $z_i$ in terms of the known $x_n$ and, since $\Delta t$ is also known to calculate the quasinormal
frequencies $\omega_i$.

\subsection{Neutral scalar perturbation}

\begin{figure}[htbp]\centering
{\subfigure[]{\label{neutral0}
\includegraphics[height=2in,width=3.0in]{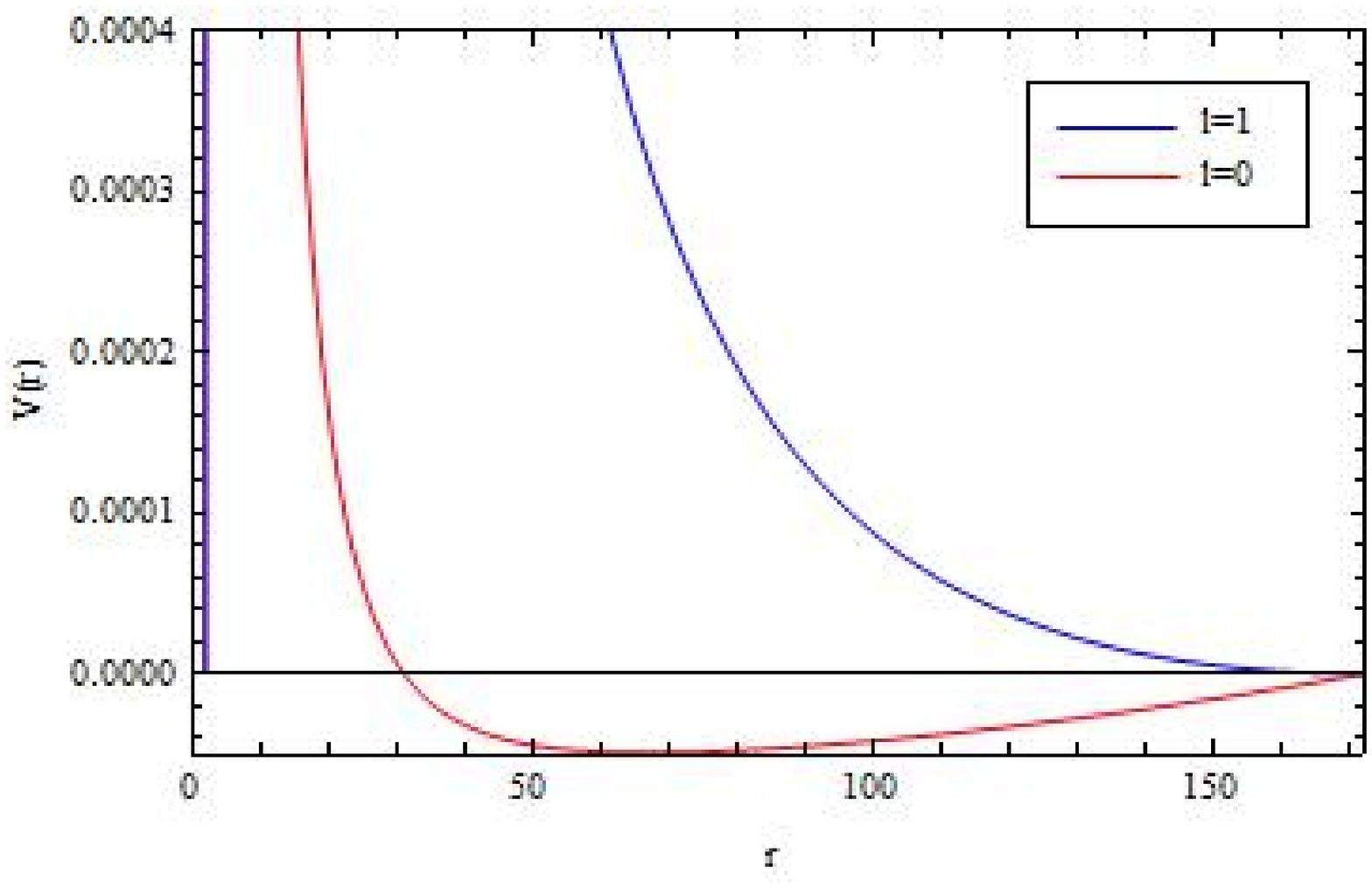}}
\subfigure[]{\label{neutral1}
\includegraphics[height=2in,width=3.0in]{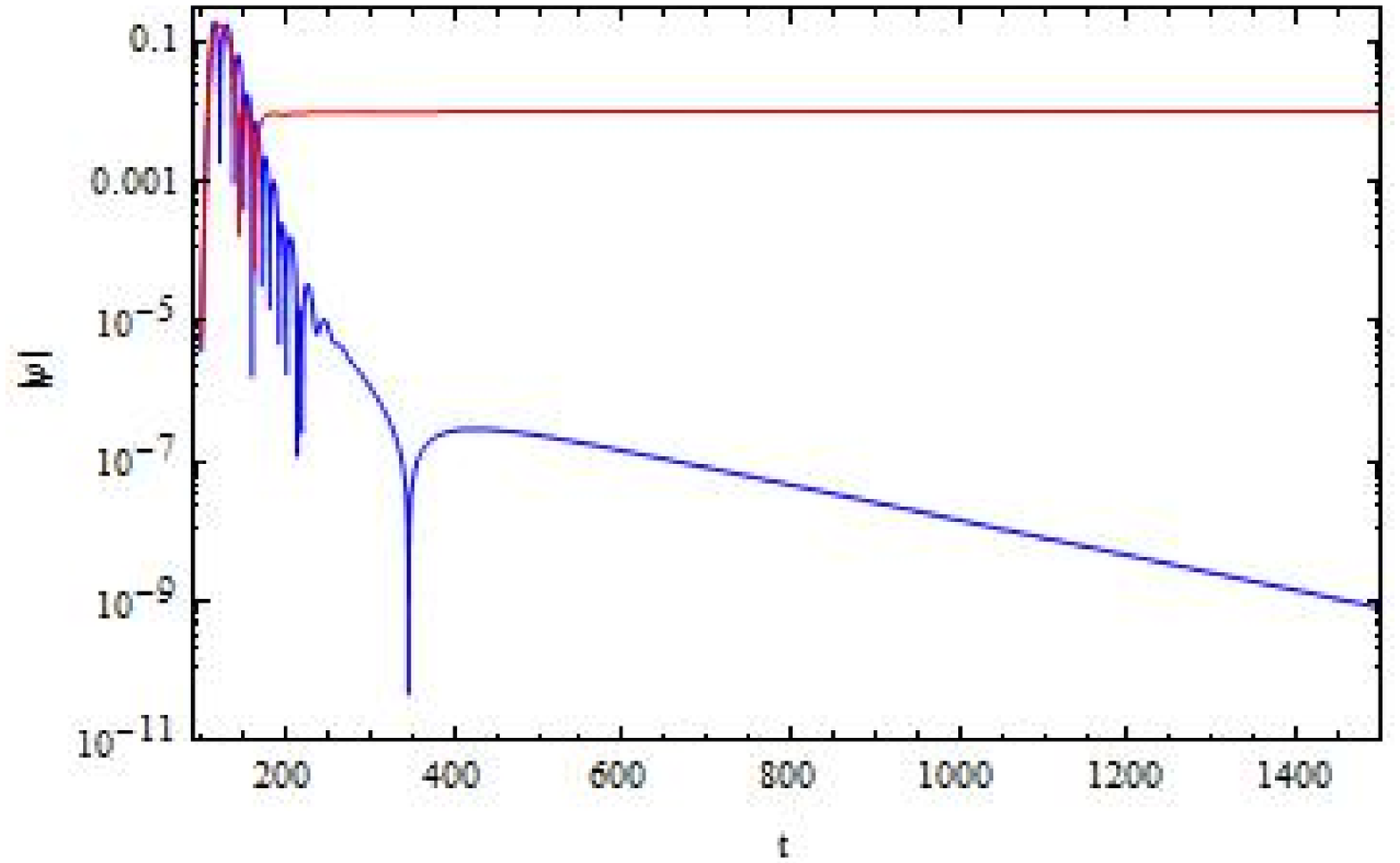}}}
\caption{(a) The potential and (b) the time-domain profiles of neutral massless scalar perturbations for $M=1$, $Q=0.5$, $\Lambda=10^{-4}$,
$l=0$ (red line) and $l=1$ (blue line). The potential $V(r)$ is obtained from the event horizon $r_+=1.866$ to the cosmological horizon
$r_c=172.197$. The field is shown outside the black hole at the position $r=100$.}\label{neutral}
\end{figure}

First, we use the finite difference method
proposed in Ref.~\cite{Abdalla:2010} to reexamine the
neutral massless scalar field perturbations in
RN-dS spacetime, which was first reported in
Ref.~\cite{Brady:1997} and later confirmed in
Ref.~\cite{Molina:2004}. For $l=0, 1$ modes, we plot
the potential in Fig.~\ref{neutral0}. We can see
that for the $l=0$ mode there exists a negative
potential well following a potential barrier near
the cosmological event horizon. But for the $l=1$
mode the potential well does not exist. The
time-domain evolution of the perturbation in
Fig.~\ref{neutral1} shows that when $l=0$, the
field rapidly settles down to a constant value
after some quasinormal oscillations. When $l=1$,
the perturbation field falls off exponentially
after the quasinormal oscillations. These results
show very good agreement with those observed in
Refs.~\cite{Brady:1997, Molina:2004}, which gives us
confidence in the numerical method we employed.
The neutral massless scalar field perturbations
do not grow in the time-domain profiles, although
there is a negative potential well when $l=0$. 
This shows that the RN-dS black hole is
stable against the neutral massless scalar
perturbation.  But when the multipole index $l=0$, the late time
tail of the neutral scalar perturbation settles down to a constant value,
instead of a decay, implying that the $l=0$ mode is prone to instability.

\subsection{Charged scalar perturbation}

It is of great interest to extend the study to
the  charged scalar field perturbation.  One
expects that with the new term in (9), the
potential will become more negative when the
scalar field is charged. Whether this can lead to a
deeper potential well and destabilize
the RN-dS black hole configuration is a question
to be answered.

\begin{figure}[htbp]\centering
{\subfigure[]{\label{charge-1}
\includegraphics[height=2in,width=3.0in]{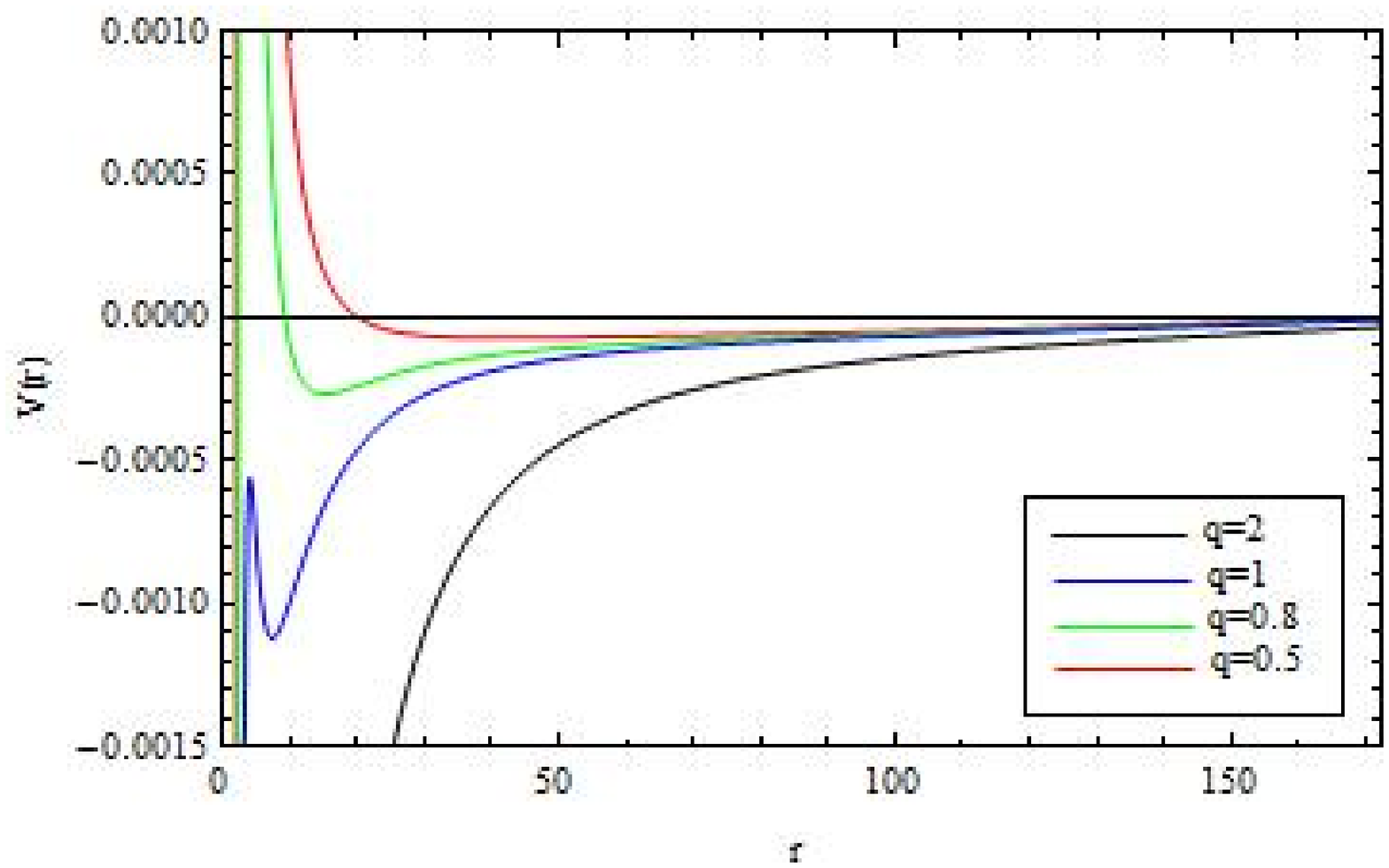}}
\subfigure[]{\label{charge-2}
\includegraphics[height=2in,width=3.0in]{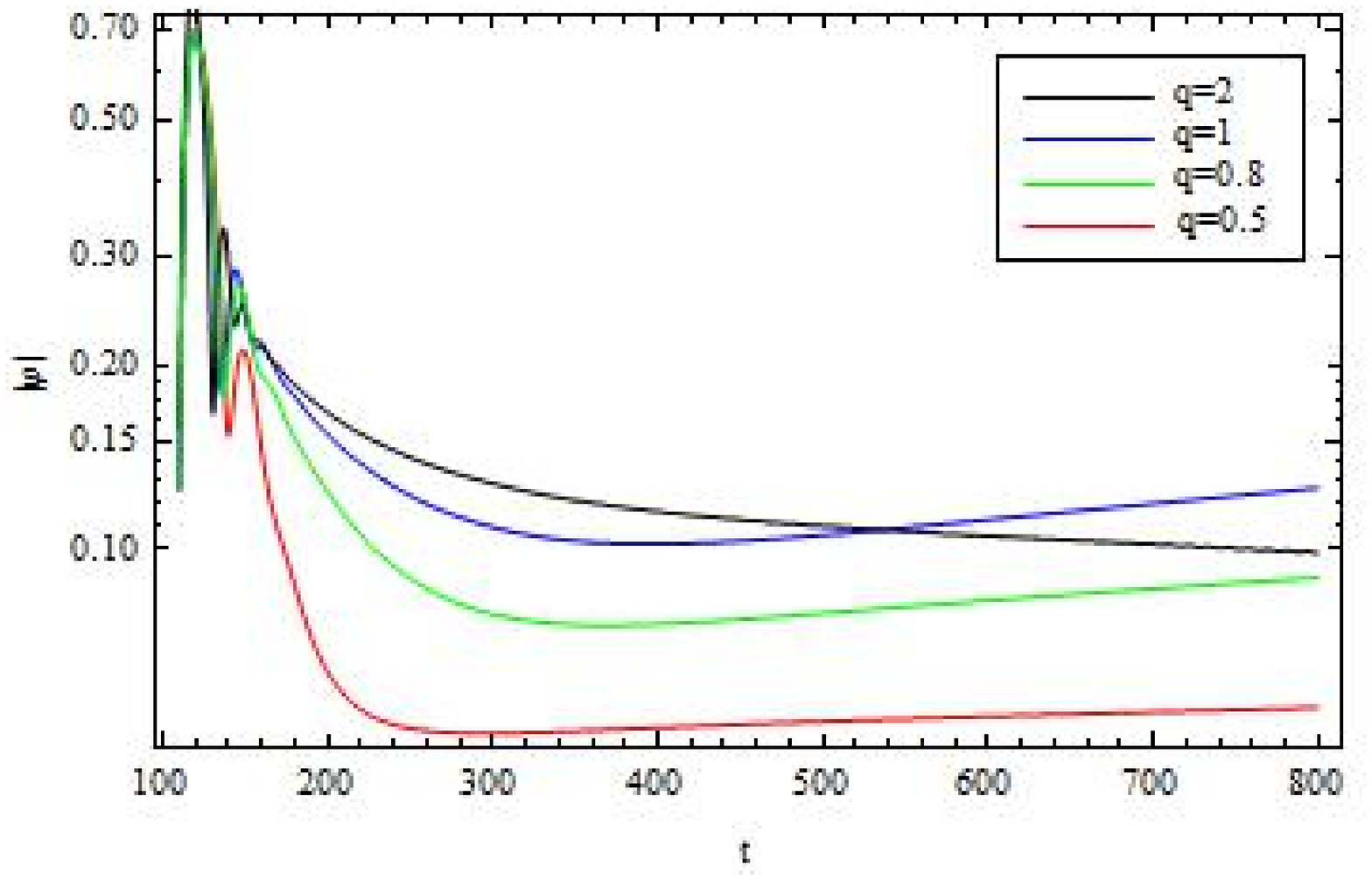}}}
\caption{(a) The potential and (b) the time-domain profiles of charged massless scalar perturbations for $M=1$, $Q=0.5$, $\Lambda=10^{-4}$,
$l=0$ and $q=0.5, 0.8, 1, 2$. The potential $V(r)$ is obtained from the event horizon $r_+=1.866$ to the cosmological horizon
$r_c=172.197$. The field is shown outside the black hole at the position $r=100$.}\label{charge1}
\end{figure}

We first concentrate on the perturbation with the
angular index $l=0$ for the charged massless
scalar field. In Fig.~\ref{charge-1} we exhibit
the potential behavior with the increase of the
charge of the massless scalar field. It is clear
that with the increase of the charge of the
scalar field, the potential has wider and deeper
negative well and the potential barrier becomes
smaller. For big enough charge $q$, i.e. $q\geq
2$, the potential barrier disappears and the
potential falls continuously from a small
negative value at the cosmological event horizon
to a very negative constant at the black hole
event horizon. Our main concern are regions where
the effective potentials contain negative values, since
possible instabilities are usually indicated in
such regions.

The response of a stable RN-dS black holes to
external charged scalar perturbation is
determined by the late time evolution of the
perturbation. In Fig.~\ref{charge-2} we show the
time-domain profiles for the evolution of charged
scalar perturbations. When the scalar field is
charged, the late time tail of the perturbation
will grow in the end. With the increase of the
charge of the scalar field, the growth will
become stronger. This can be determined by the
slopes of the late time tails. The growth of the
charged scalar perturbation can be attributed to
the negative potential well, which can trap and
accumulate the perturbation and finally destabilize the background RN-dS black hole
spacetime. But when the charge of the scalar
field is large enough, its tail decays instead of
growing in the end. This can be understood from
the corresponding monotonic behavior in the
potential, which cannot hinder the perturbation
from falling into the black hole.

\begin{figure}[htbp]
\centering
\includegraphics[width=4.0in]{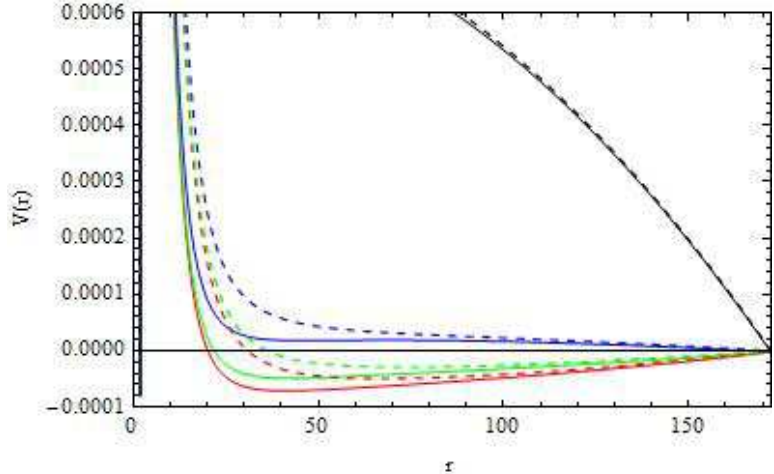}
\caption{The potential $V(r)$ is described from the event horizon $r_+=1.866$ to the cosmological horizon
$r_c=172.197$ with $M=1$, $Q=0.5$, $\Lambda=10^{-4}$, and $l=0$. The charge of the scalar field are $q=0$ (dashed) and $q=0.5$ (solid).
The masses of the field are $\mu=0$ (red line), $\mu=0.005$ (green line), $\mu=0.01$ (blue line), and $\mu=0.03$ (black line) from bottom to top.}\label{mass-1}
\end{figure}

Above, we focused on the massless charged scalar
field. It is of interest to explore the influence
of the scalar field mass on the stability of
the black hole. For the massive charged scalar
field,  we observe the influence on the potential
by different values of the mass of the scalar
field in Fig.~\ref{mass-1}.  The dashed lines are
for the neutral massive scalar fields, while the
solid lines are for the charged scalar fields
with different masses. We see that, when the
scalar fields become more massive, the potential
wells become narrower and less negative. When the
mass of the scalar field is large enough, the
potential will have only a positive barrier
between the black hole and cosmological event
horizons.

\begin{figure}[htbp]\centering
{\subfigure[]{\label{mass1}
\includegraphics[height=2in,width=3.0in]{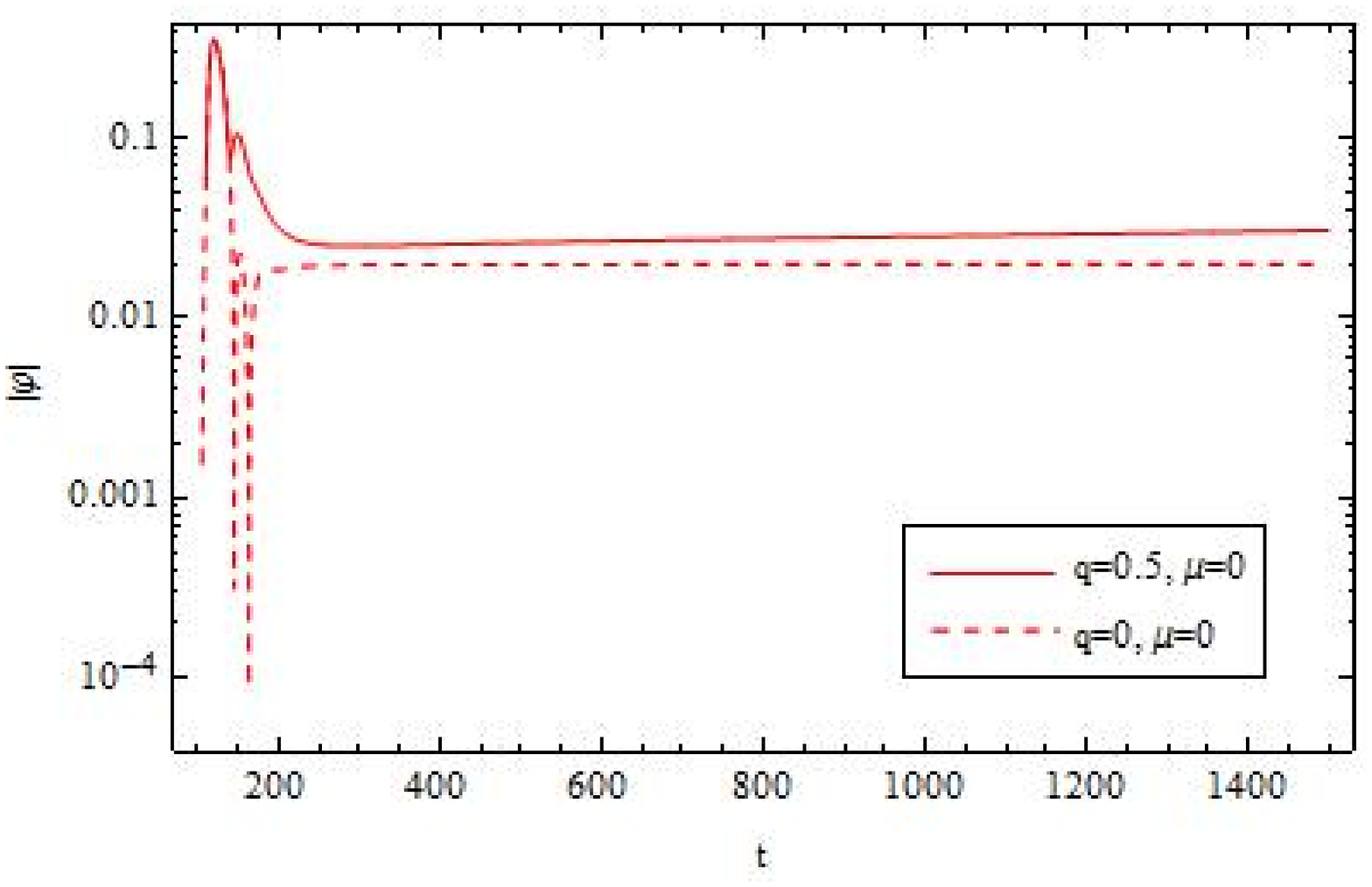}}
\subfigure[]{\label{mass2}
\includegraphics[height=2in,width=3.0in]{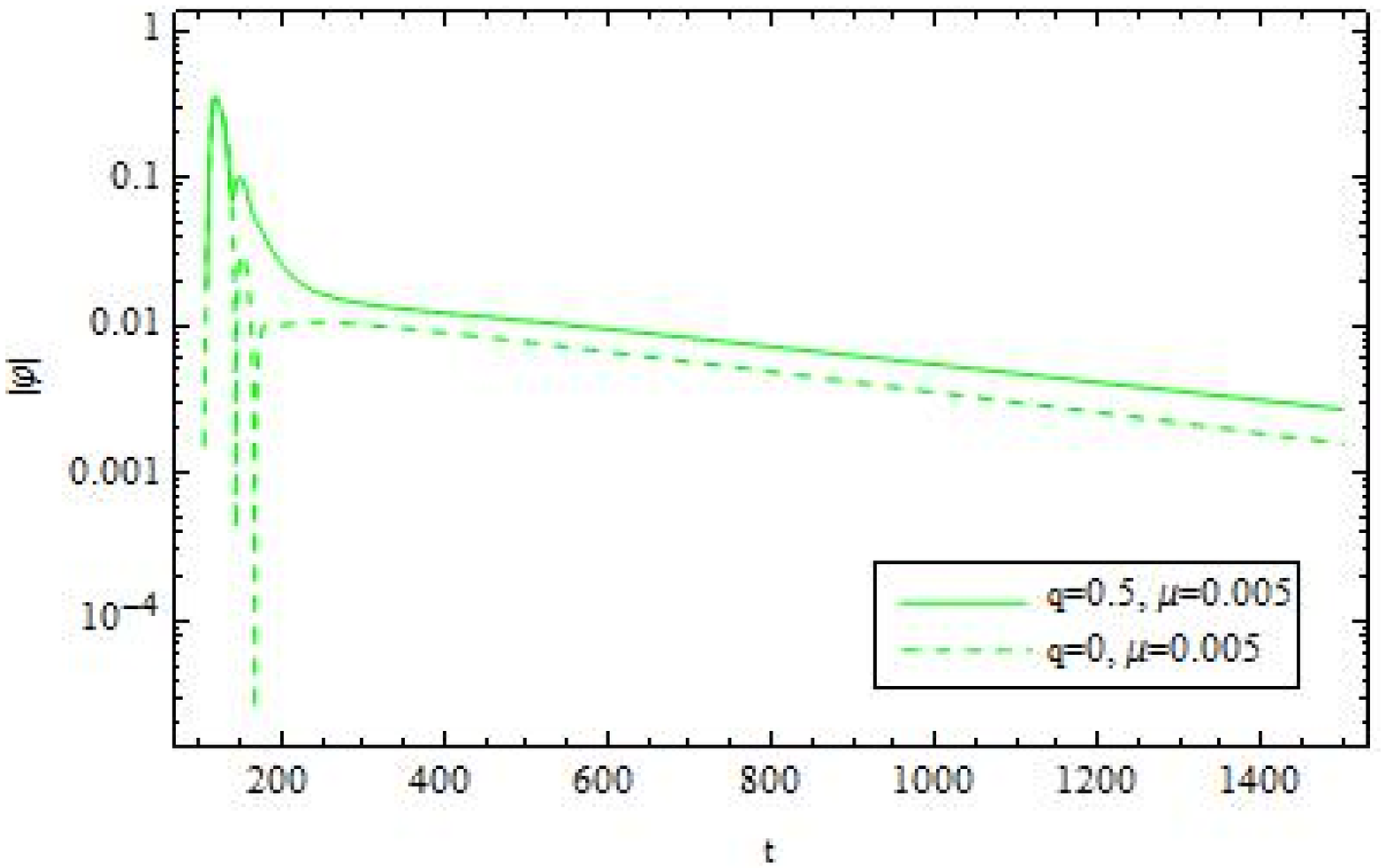}}
\subfigure[]{\label{mass3}
\includegraphics[height=2in,width=3.0in]{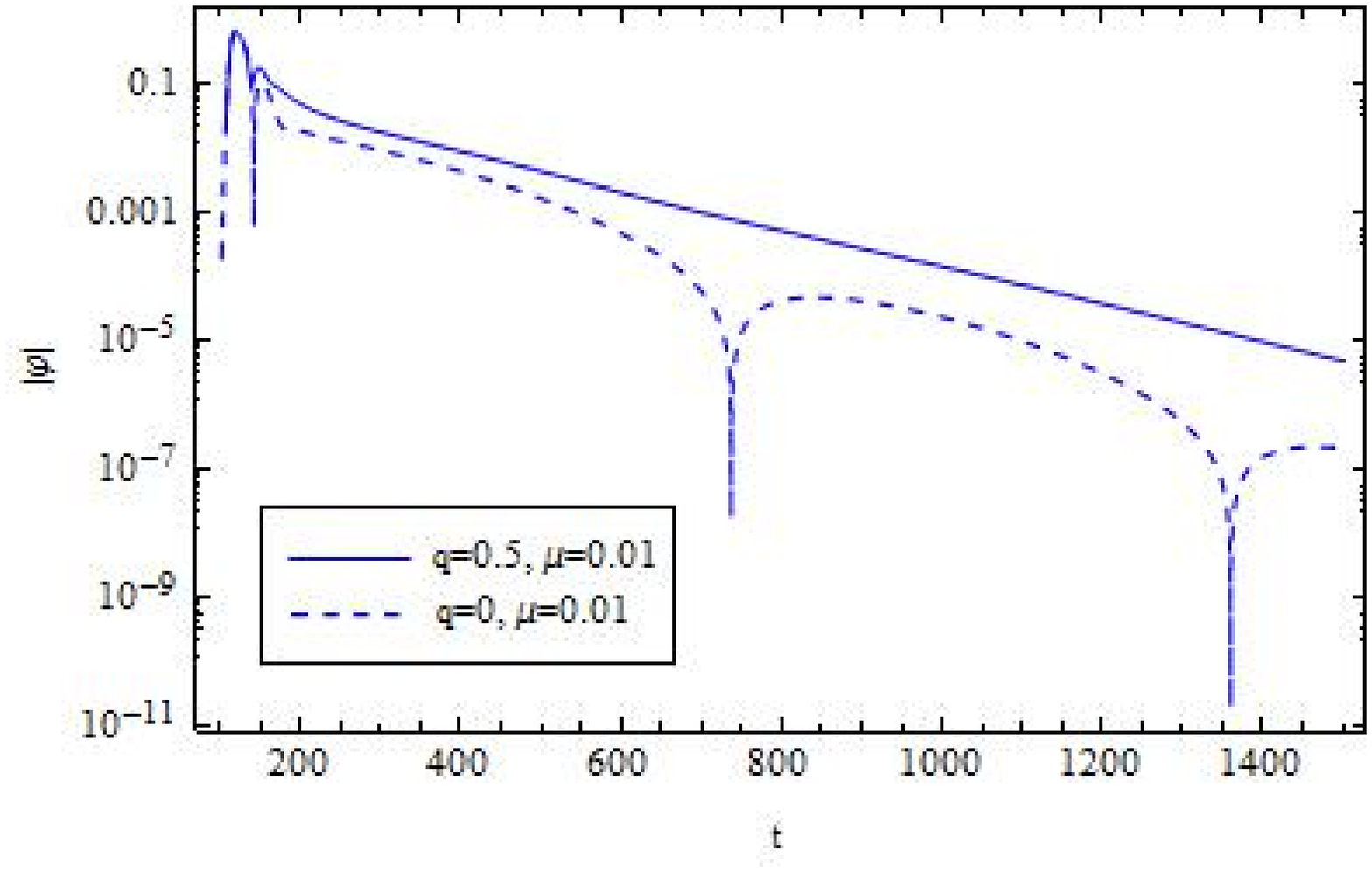}}
\subfigure[]{\label{mass4}
\includegraphics[height=2in,width=3.0in]{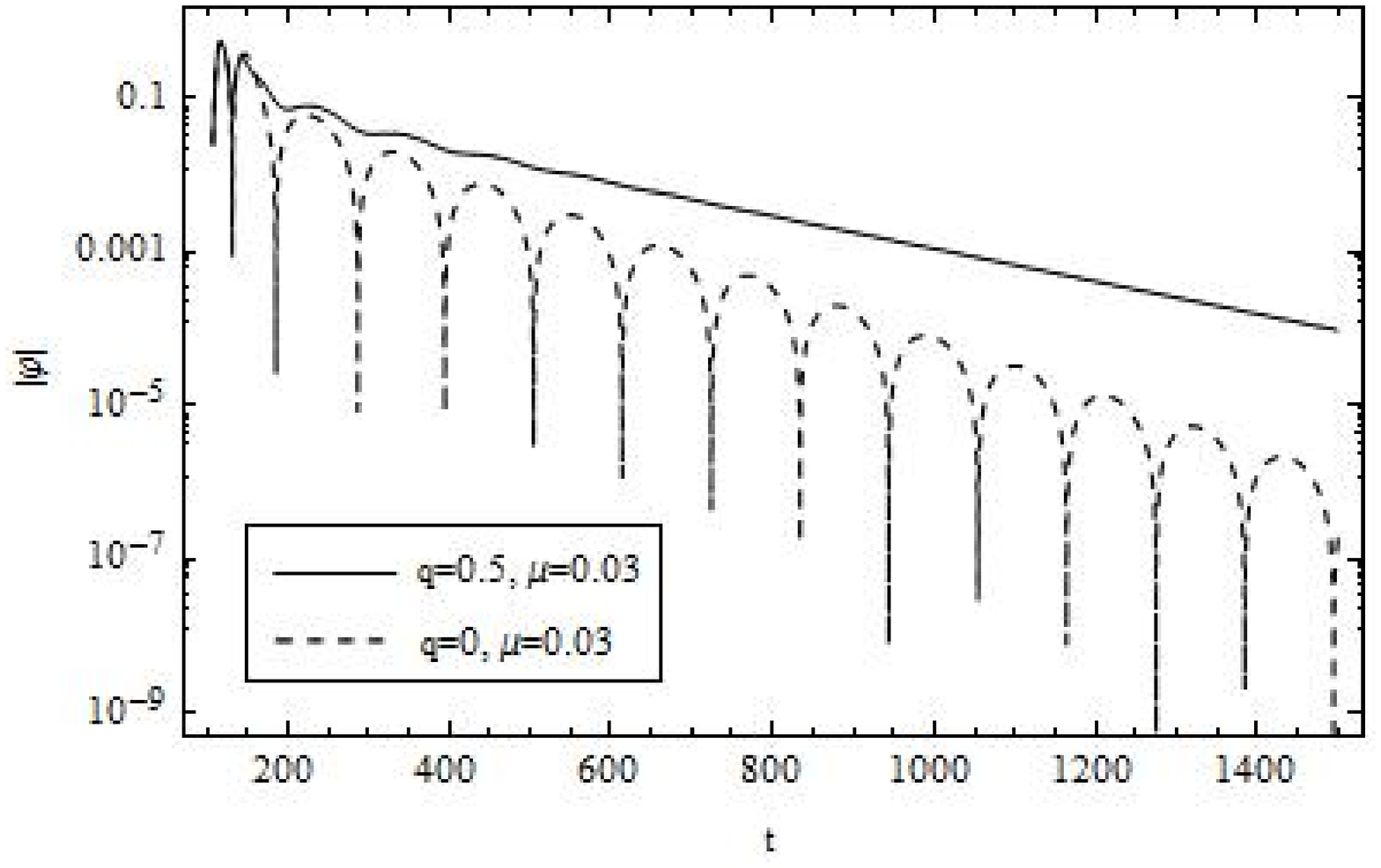}}}
\caption{Time-domain profiles for the scalar perturbations at $r=100$ with $M=1$, $Q=0.5$, $\Lambda=10^{-4}$, and
$l=0$. The solid lines and the dashed lines denote $q=0.5$ and $q=0$, respectively. The masses of the field are
(a) $\mu=0$, (b) $\mu=0.005$, (c) $\mu=0.01$, and (d) $\mu=0.03$. }\label{mass-2}
\end{figure}

The evolution of perturbations is shown in
Fig.~\ref{mass-2}. The dashed lines are for the
neutral scalar field.  We see that only for the
massless scalar perturbation, which contains a
constant tail, all massive perturbations exhibit
a decay behavior at late time. When the mass of
the scalar field is large enough, i.e.
$\mu\geq0.01$, neutral scalar perturbations have
a long-lived ringing at the late time. The solid
lines are for the scalar fields with fixed charge
$q=0.5$ but different masses. For the massless
case, it exhibits a growing mode, but when the
scalar field becomes massive, the perturbation
decays. Fitting the curves in Fig.~\ref{mass-2}
with function $|\varphi|\propto e^{\kappa t}$ at
late time, we can get the decay rate of
perturbations in Table \ref{table-1}. It is shown
that, for both the neutral and charged scalar
field perturbations, the decay of tails first
becomes quicker, and then slows down with the
increase of the mass of the scalar field. The
massive field  exhibiting oscillations in the
late time tail has higher energy, so that it
decays slower. The influence of the scalar mass
on the behavior of decay in the late time tail is
consistent with that reported in the Kerr-Newman
background \cite{Konoplya:2013}.

In summary we find that for the $l=0$ mode, when the
scalar field is charged, the perturbation can grow. But such growth can be avoided
if we increase the mass of the scalar field.
The  RN-dS black hole background
can be destabilized by the charged scalar
field perturbation.


\begin{table}[H]
\begin{center}
\begin{tabular}{|c|c|c|c|c|}
\hline
& $\mu=0$ & $\mu=0.0005$ & $\mu=0.01$ & $\mu=0.03$ \\
\hline
$q=0$ & $0$ & $-0.00158$ & $-0.009$ & $-0.0083$ \\
\hline
$q=0.5$ & $0.000153$ & $-0.0014$ & $-0.0067$ & $-0.0049$ \\
\hline
\end{tabular}
\end{center}
\caption{The decay rate of the scalar perturbations fit $\propto e^{\kappa t}$ at late time
for $l=0$, $M=1$, $Q=0.5$ and $\Lambda=10^{-4}$. Here we obtained the value of $\kappa$ for different value of the charge and mass of the field.}\label{table-1}
\end{table}

\begin{figure}[htbp]\centering
{\subfigure[]{\label{charge2-1}
\includegraphics[height=2in,width=3.0in]{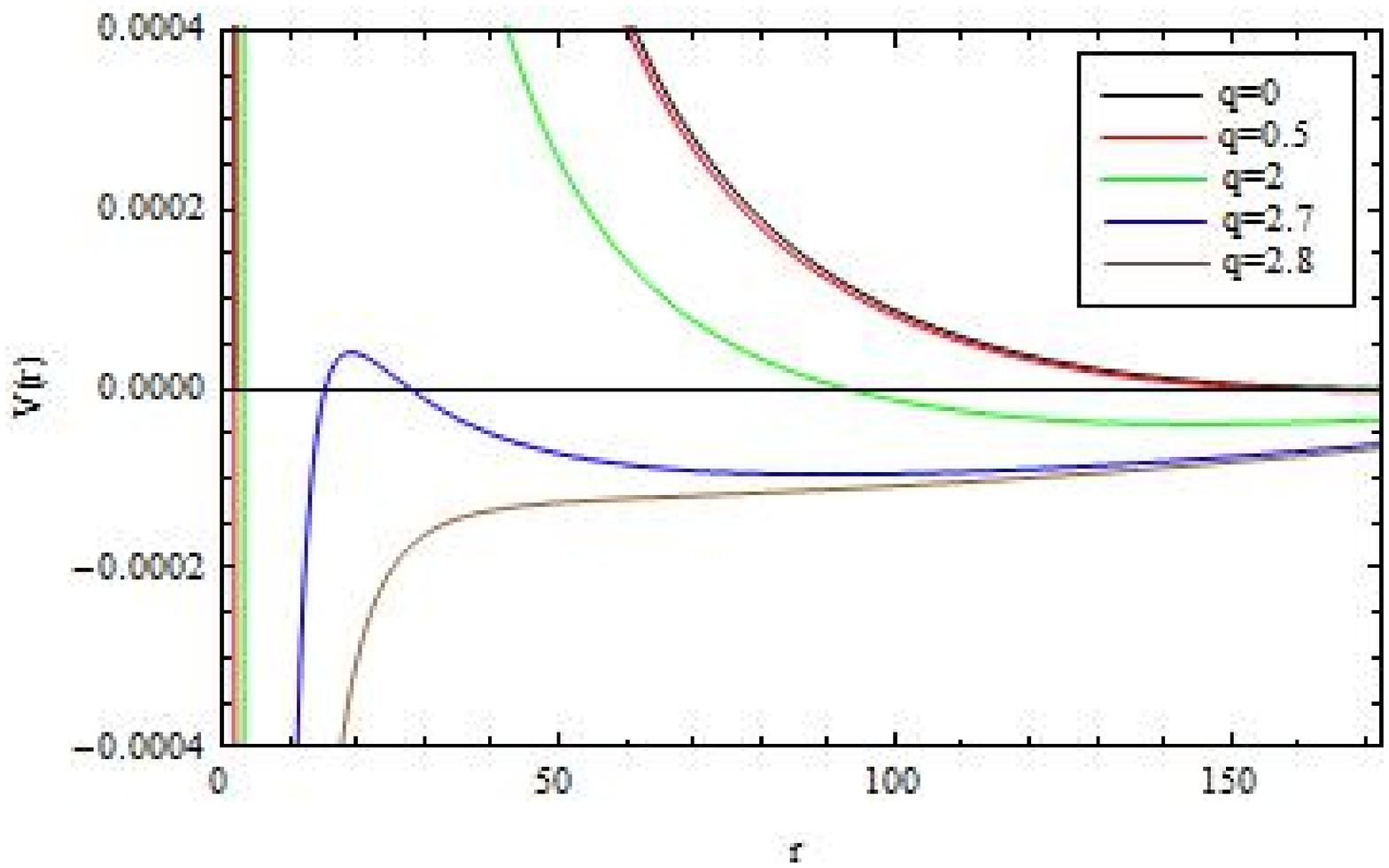}}
\subfigure[]{\label{charge2-2}
\includegraphics[height=2in,width=3.0in]{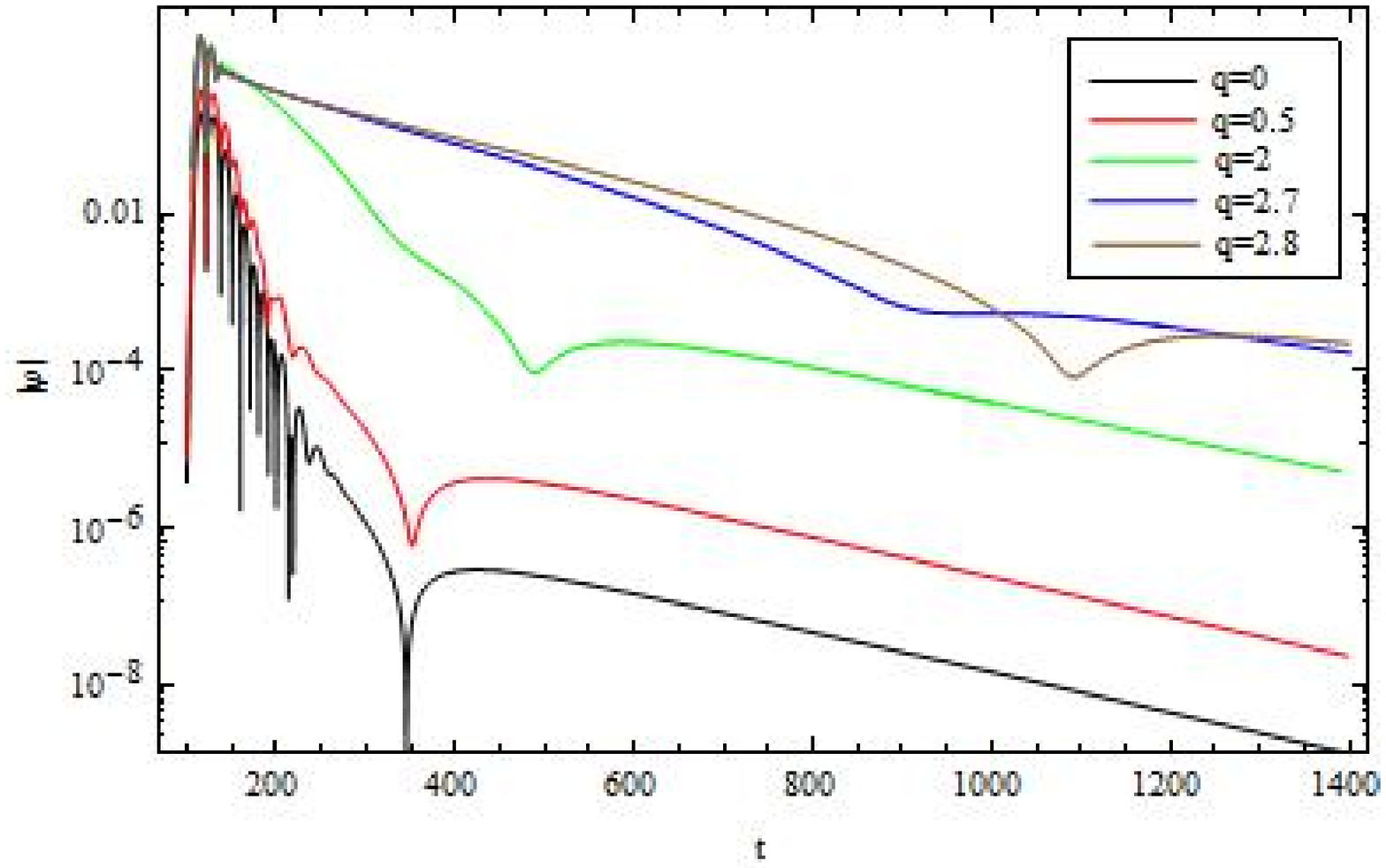}}}
\caption{(a) The potential and (b) the time-domain profiles of charged massless scalar perturbations for $M=1$, $Q=0.5$, $\Lambda=10^{-4}$,
$l=1$, and $q=0, 0.5, 2, 2.7, 2.8$. The potential $V(r)$ is obtained from the event horizon $r_+=1.866$ to the cosmological horizon
$r_c=172.197$. The field is shown outside the black hole at the position $r=100$.}\label{charge2}
\end{figure}

Now, we discuss the $l=1$ perturbation
mode. We first look at the massless but charged
scalar field. In Fig.~\ref{charge2-1}, we plotted
the potential with the change of the charge $q$.
It is similar to the one we observed for the $l=0$ mode.
With the increase of the charge, the region of
negative potential becomes wider and deeper,
while the potential barrier becomes lower. When
the charge of the field is over a certain critical value,
the potential becomes negative everywhere and
decreases monotonically from the cosmological
horizon to the black hole event horizon.

In contrast to the situation in the $l=0$ mode,
the potentials with negative wells here do not
imply instability.  The time-domain profiles of
perturbation are shown in Fig.~\ref{charge2-2}.
We see that, when the charge of the scalar field
increases, the decay of the perturbation tails
becomes slower.We calculated that even until $q=5$, which is
$10$ times the charge of the black hole itself,
we still have the decay behavior of the
perturbation. For larger $q$, in the numerical
computation we require smaller $\Delta r_*$ to
ensure convergence, what takes a much longer
computation time.  Thus, in contrast with the $l=0$
case, the RN-dS black hole background remains stable
against charged scalar perturbations
when the angular index is $l=1$. This result
supports the argument given in Ref.~\cite{Bronnikov:2012},
that the negative effective potential does not guarantee any growing mode  in
the time-domain profiles for the evolution of the
perturbations. The negative potential well can be
viewed as a necessary condition for the
instability of the black hole, rather than a
sufficient condition. Generally, a bound state structure is required
for the equivalent Schr\"odinger problem \cite{bachelotbachelot}.

\begin{figure}[htbp]\centering
{\subfigure[]{\label{mass2-3}
\includegraphics[height=2in,width=3.0in]{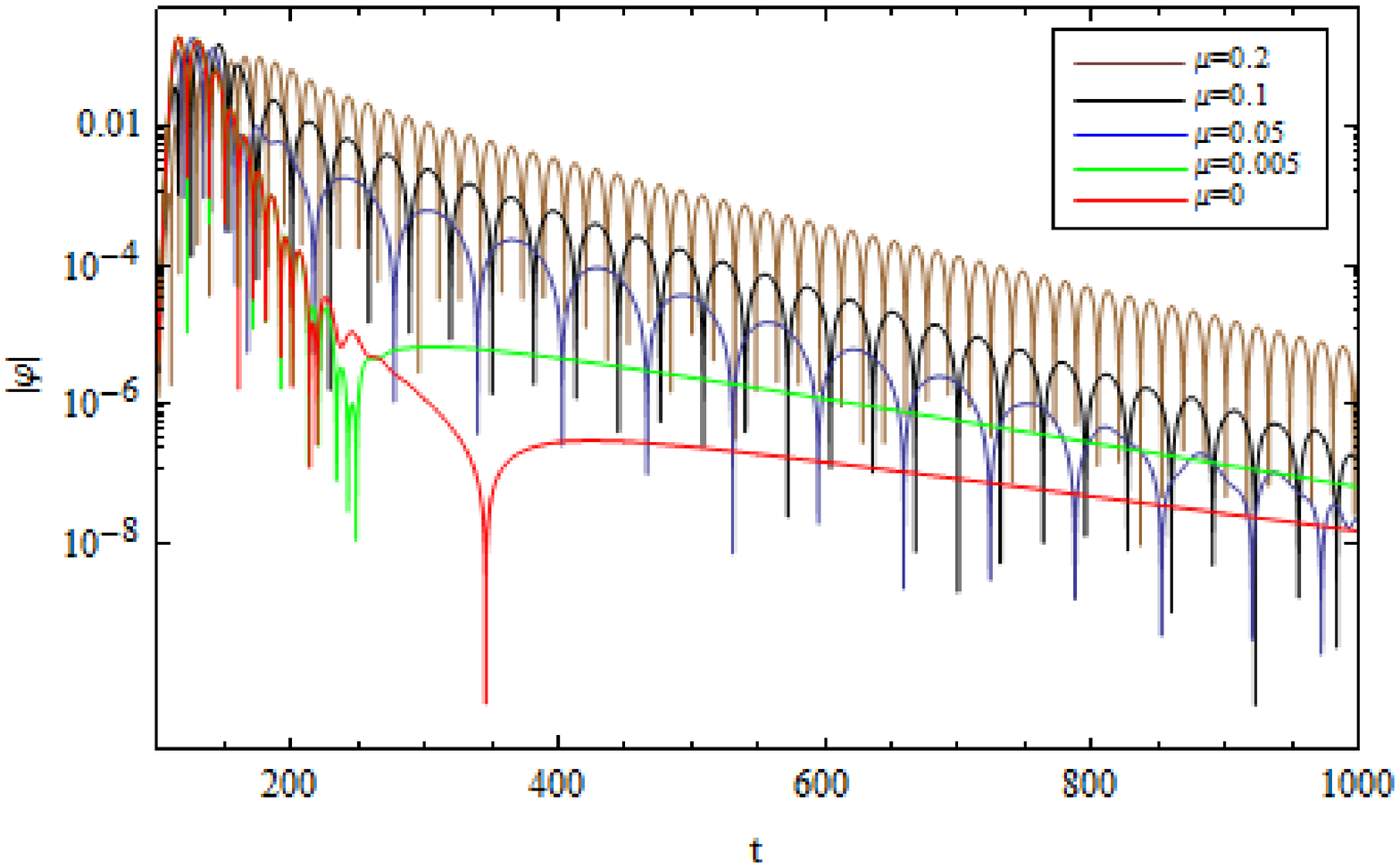}}
\subfigure[]{\label{mass2-4}
\includegraphics[height=2in,width=3.0in]{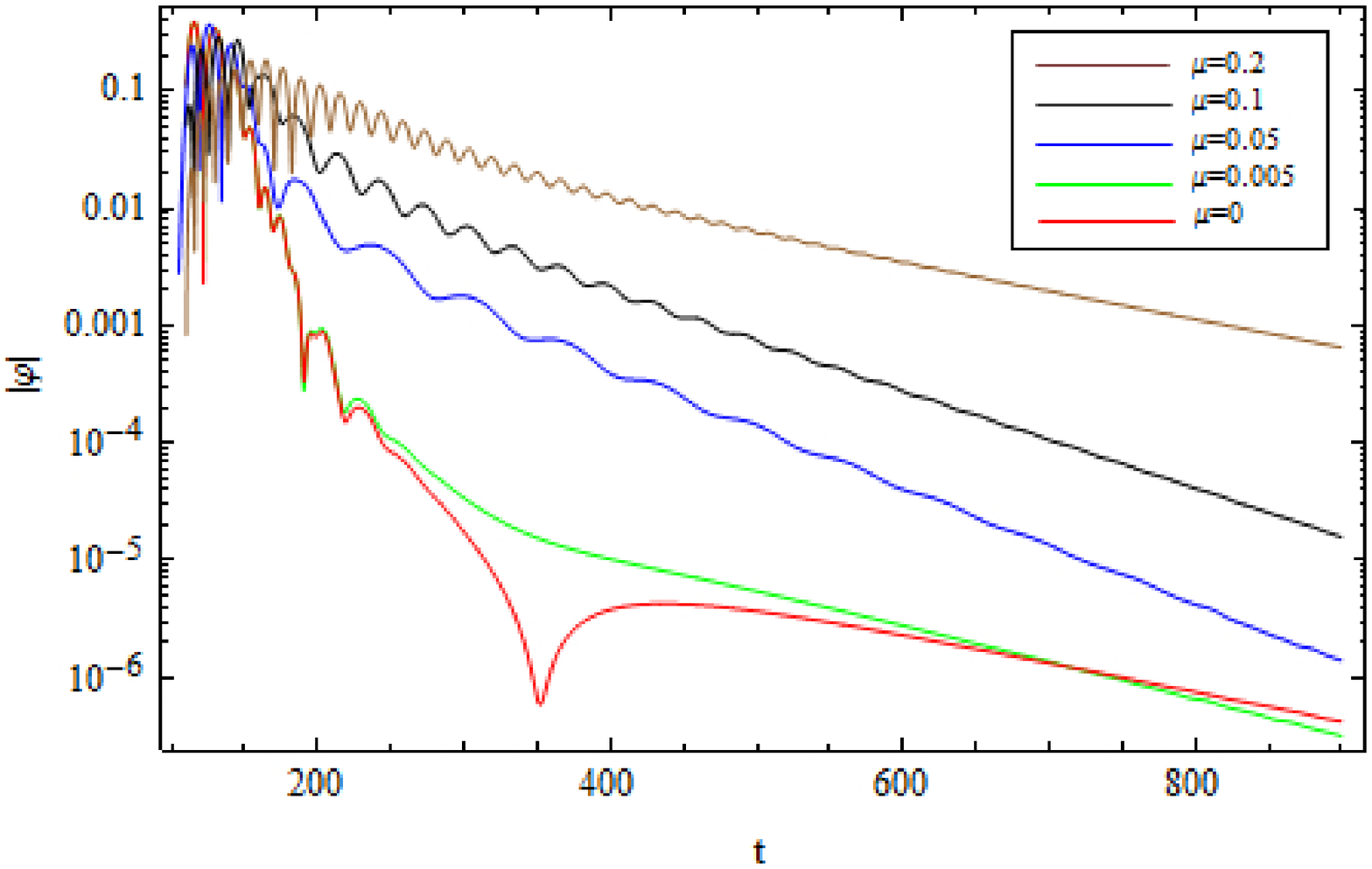}}}
\caption{Time-domain profiles for the scalar perturbations at $r=100$ with $M=1$, $Q=0.5$, $\Lambda=10^{-4}$,
and $l=1$. We consider (a) a neutral field and (b) a charged field with $q=0.5$. The masses of the field are
$\mu=0$ (red line), $\mu=0.005$ (green line), $\mu=0.05$ (blue line), $\mu=0.1$ (black line), and $\mu=0.2$ (brown line). }\label{massl1}
\end{figure}

For the $l=1$ mode, the influence of the mass of the
scalar field on the perturbation is illustrated
in Fig.~\ref{massl1} for both neutral and charged
scalar fields with fixed $q=0.5$. It is shown that
with the increase of the mass of the scalar
field, no matter whether it is neutral or charged, the
decay of perturbations first becomes quicker,
and then slows down. This result can be seen in
Table \ref{table-2} with the decay rate of
perturbations fitted by $|\varphi|\propto
e^{\kappa t}$ at late time, and is consistent
with the $l=0$ case. When the mass of the scalar
field is large enough, say $\mu\geq0.05$, one can
see that the late time tail of a neutral scalar
perturbation is dominated by a long-lived
ringing. But for the charged scalar field,  the
ringing will be followed by an exponential tail.

\begin{table}[H]
\begin{center}
\begin{tabular}{|c|c|c|c|c|c|}
\hline
& $\mu=0$ & $\mu=0.0005$ & $\mu=0.05$ & $\mu=0.1$ & $\mu=0.2$ \\
\hline
$q=0$ & $-0.00577$ & $-0.00736$ & $-0.0135$ & $-0.0129$ & $-0.0105$ \\
\hline
$q=0.5$ & $-0.00566$ & $-0.00724$ & $-0.0114$ & $-0.00956$ & $-0.00551$ \\
\hline
\end{tabular}
\end{center}
\caption{The decay rate of the scalar perturbations fit $\propto e^{\kappa t}$ at late time
for $l=1$, $M=1$, $Q=0.5$ and $\Lambda=10^{-4}$. Here we obtained the value of $\kappa$ for different value of the charge and mass of the field.}\label{table-2}
\end{table}

\section{discussion}
The discussions presented in the previous section
are focused on numerical calculations, but do not
offer much physical insight.   Could the
instability of the charged scalar field
perturbation for the $l=0$ state be triggered by the
superradiant instability? If we look at the
effective potential, there is a valley following
the barrier. This is similar to the
characteristic potential to accommodate the
superradiant instability discussed in
Refs.~\cite{Hod:2012,Hod:2013, Zhang:2013}.  If there
is superradiance, in the valley of the potential
the reflection from the potential barrier could
be amplified and its energy could be accumulated
continuously which could lead to instability.

For the RN-dS case, to examine the superradiant instability, we have to consider the classical
scattering problem for the charged scalar field.  The problem can be reduced to the wavelike equation
\begin{eqnarray}
{{d^2\Psi}\/{dr_*^2}}+(\omega^2-\overline{V}(r))\Psi=0\;,\label{radial}
\end{eqnarray}
with the effective potential
\begin{eqnarray}
\overline{V}(r)={2qQ\omega\/r}-{q^2Q^2\/{r^2}}+f(r)\bigg({l(l+1)\/{r^2}}+\mu^2+{f'(r)\/r}\bigg)\;.\label{potential2}
\end{eqnarray}
The effective potential has the following asymptotic behavior: when $r\rightarrow r_+$, $\overline{V}\approx {2qQ\omega\/r_+}-{q^2Q^2\/{r_+^2}}$ and when $r\rightarrow r_c$,
$\overline{V}\approx{2qQ\omega\/r_c}-{q^2Q^2\/{r_c^2}}$ respectively.
In a scattering experiment, (\ref{radial}) has the following asymptotic behavior
\begin{eqnarray}
\Psi\sim Be^{-i(\omega-qQ/r_+)r_*}\;, \, \, \, as\, \, r\rightarrow r_+\,(r_*\rightarrow-\infty)\;,\label{ingoing}
\end{eqnarray}
\begin{eqnarray}
\Psi\sim e^{-i(\omega-qQ/r_c)r_*}+A e^{+i(\omega-qQ/r_c)r_*}\;, \, \, \, as\, \, r\rightarrow r_c\,(r_*\rightarrow\infty)\;,
\end{eqnarray}
where $A$ is called the amplitude of the reflected wave or the reflection coefficient, and $B$ is the transmission coefficient.
We adopted the boundary condition that the wave comes from the cosmological horizon, partially passes through
the potential barrier and falls inside the event horizon, while the rest reflects back to the cosmological horizon
\cite{Khanal:1985,Tachizawa:1993}. Using the constancy of the Wronskian, we can show that
\begin{eqnarray}
1-|A|^2={{\omega-qQ/r_+}\/{\omega-qQ/r_c}}|B|^2\;.
\end{eqnarray}
The reflected wave has larger amplitude than the incident
one when
\begin{eqnarray}
{qQ\/r_c}<\omega<{qQ\/r_+}\;.\label{regime}
\end{eqnarray}
Such an amplification of the incident wave is called
superradiance. When the cosmological constant $\Lambda$ approaches to zero,
$r_c$ becomes infinite, and the superradiant condition Eq.~(\ref{regime}) is equivalent to the one for the
RN black hole in the presence of charged scalar perturbations, $0<\omega<{qQ/r_+}$ \cite{Bekenstein:1973,Hod:2012,Hod:2013}.
Here, $\omega$ is the real oscillation frequency of the perturbation.

Using the Prony method to fit the data in
Fig.~\ref{charge-2} at late time, we can obtain
the dominant frequencies of the modes in Table
\ref{table-3} for charged scalar perturbations
and vanishing angular momentum, $l=0$. One can
see that both the growing modes and decay mode
satisfy the superradiant condition
(\ref{regime}). This means that the instability
that we found is caused by the superradiance, but
not all the superradiant modes are unstable. This
supports the discussion in a recent work
\cite{Konaplya:2014} where it was argued
numerically and analytically that the
superradiance is a necessary condition, instead
of a sufficient condition, for the instability.

\begin{table}[H]
\begin{center}
\begin{tabular}{|c|c|c|c|c|c|}
\hline
$q$ & $\omega$ & ${qQ/r_c}$ & ${qQ/r_+}$ \\
\hline
$0.5$ & $0.001508+0.000153i$ & $0.001452$ & $0.133958$ \\
\hline
$0.8$ & $0.002559+0.000392i$ & $0.002323$ & $0.214333$ \\
\hline
$1$   & $0.003414+0.000557i$ & $0.002904$ & $0.267916$ \\
\hline
$2$   & $0.007487-0.000331i$ & $0.005807$ & $0.535831$ \\
\hline
\end{tabular}
\end{center}
\caption{Dominant modes for $M=1$, $Q=0.5$, $\Lambda=10^{-4}$ and $l=0$.}\label{table-3}
\end{table}

\section{conclusion}
In this paper, we have investigated the  stability of the RN-dS black hole. In the four-dimensional spacetime, the RN-dS
black hole was found to be stable against the neutral scalar field perturbation \cite{Brady:1997, Brady:1999, Molina:2004}.
However, recently, it has been reported that, when the spacetime dimensionality is $D > 6$,  the RN-dS black hole can
become unstable if there are small gravitational perturbations of scalar type \cite{Konoplya:2009, Cardoso:2010}.
Here we found that, even in the four-dimensional spacetime, the instability of RN-dS black hole can appear under the
perturbation of charged scalar field when the  angular index vanishes.  But when the angular index is unit, the
background four-dimensional RN-dS black hole remains stable.  We have noticed that the negative effective potential is
not the only criterion to decide the stability of the background configuration; the practical tool for testing
stability in spacetime backgrounds is the numerical investigation of the time-domain profiles of the perturbations.
This supported the argument in Ref.~\cite{Bronnikov:2012} and the necessity of a bound state in the equivalent Schr\"{o}dinger
problem \cite{bachelotbachelot}. We have further explored the physical nature about why the instability is triggered by the
$l=0$ mode in the four-dimensional RN-dS black hole. We found that this instability is caused by superradiance,
but not all the superradiant modes are unstable.

\begin{acknowledgments}
This work was supported in part by the National Natural Science
Foundation of China. Z. Z was also supported by China Postdoctoral Science Foundation under Grants No.
2011M500764 and No. 2012T50414. E.A. and C.E.P. thank FAPESP and CNPq (Brazil) for support.
\end{acknowledgments}


\begin{thebibliography}{99}

\bibitem{Konoplya:2011} R. A. Konoplya and A. Zhidenko, Rev. Mod. Phys. {\bf83}, 793 (2011) [arXiv:1102.4014].

\bibitem{Wang:2005} B. Wang, Braz.J.Phys. {\bf35}, 1029 (2005) [arXiv:gr-qc/0511133].

\bibitem{Konoplya:2011:295} R. Gregory and  R. Laflamme, Phys. Rev. Lett. {\bf70}, 2837 (1993) [arXiv:hep-th/9301052].

\bibitem{Konoplya:2011:296} R. Gregory and  R. Laflamme, Nucl. Phys. B {\bf428}, 399 (1994) [arXiv:hep-th/9404071].

\bibitem{Konoplya:2011:140} R. A. Konoplya and A. Zhidenko, Phys. Rev. D {\bf77}, 104004 (2008) [arXiv:0802.0267].

\bibitem{Konoplya:2011:299} M. Beroiz, G. Dotti and R. J. Gleiser, Phys. Rev. D {\bf76}, 024012 (2007) [arXiv:hep-th/0703074].

\bibitem{Konoplya:2009} R. A. Konoplya and A. Zhidenko, Phys.Rev.Lett. {\bf103}, 161101 (2009) [arXiv:0809.2822].

\bibitem{Cardoso:2010} V. Cardoso, M. Lemos and M. Marques, Phys. Rev. D {\bf80}, 127502 (2009) [arXiv:1001.0019].

\bibitem{Cardoso:2013:arXiv:1307.0038} V. Cardoso, arXiv:1307.0038.

\bibitem{Hod:2012} S. Hod, Phys. Lett. B {\bf713}, 505 (2012).

\bibitem{Hod:2013} S. Hod, Phys. Lett. B {\bf718}, 1489 (2013) [arXiv:1304.6474].

\bibitem{Cardoso:2013:50} J. C. Degollado, C. A. R. Herdeiro and H. F. R\'{u}narsson, Phys. Rev. D {\bf88}, 063003 (2013) [arXiv:1305.5513].

\bibitem{Hod:2013-2} S. Hod, Phys. Rev. D {\bf88}, 064055 (2013) [arXiv:1310.6101].

\bibitem{Zhang:2013} S. -J. Zhang, B. Wang and E. Abdalla, arXiv:1306.0932.

\bibitem{arXiv:1002.2679} X. He, B. Wang, R. -G. Cai and C. -Y. Lin, Phys. Lett. B {\bf688}, 230 (2010) [arXiv:1002.2679].

\bibitem{Abdalla:2010} E. Abdalla, C. E. Pellicer, J. de Oliveira and A. B. Pavan, Phys. Rev. D {\bf82}, 124033 (2010) [arXiv:1010.2806].

\bibitem{1111.6729} Y. Liu and B. Wang, Phys. Rev. D {\bf85}, 046011 (2012) [arXiv:1111.6729].

\bibitem{Brady:1997} P. R. Brady, C. M. Chambers, W. Krivan and P. Laguna, Phys. Rev. D {\bf55}, 7538 (1997) [arXiv:gr-qc/9611056].

\bibitem{Brady:1999} P. R. Brady, C. M. Chambers, W. G. Laarakkers and E. Poisson, Phys. Rev. D {\bf60}, 064003 (1999) [arXiv:gr-qc/9902010].

\bibitem{Molina:2004} C. Molina, D. Giugno, E. Abdalla and A. Saa, Phys. Rev. D {\bf69}, 104013 (2004) [arXiv:gr-qc/0309079].

\bibitem{Ching:1995PRL} E. S. C. Ching, P. T. Leung, W. M. Suen and K. Young, Phys. Rev. Lett. {\bf74}, 4588 (1995) [arXiv:gr-qc/9408043].

\bibitem{Bronnikov:2012} K. A. Bronnikov, R. A. Konoplya and A. Zhidenko, Phys. Rev. D {\bf86}, 024028 (2012) [arXiv:1205.2224].

\bibitem{bachelotbachelot} A. Bachelot and A. Motet-Bachelot, Ann. I. H. P.: Phys. Theor. {\bf 59}, 3 (1993).

\bibitem{Wang:2004} B. Wang, C. -Y. Lin and C. Molina, Phys. Rev. D {\bf70}, 064025 (2004) [arXiv:hep-tp/0407024].

\bibitem{Konoplya:2013} R. A. Konoplya and A. Zhidenko, Phys. Rev. D {\bf88}, 024054 (2013) [arXiv:1307.1812].

\bibitem{Berti:2007} E. Berti, V. Cardoso, J. A. Gonz\'{a}lez and U. Sperhake, Phys. Rev. D {\bf75}, 124017 (2007) [arXiv:gr-qc/0701086].

\bibitem{Khanal:1985} U. Khanal, Phys. Rev. D {\bf32}, 879 (1985).

\bibitem{Tachizawa:1993} T. Tachizawa and K. Maeda, Phys. Lett. A {\bf172}, 325 (1993).

\bibitem{Bekenstein:1973} J. D. Bekenstein, Phys. Rev. D {\bf7}, 949 (1973).

\bibitem{Konaplya:2014} R. A. Konoplya and A. Zhidenko, arXiv:1406.0019.


\end{thebibliography}
\end{document}